\documentclass[pra,aps,showpacs,twocolumn,floatfix,10pt]{revtex4-1}
\usepackage{amsmath,amsfonts,amssymb,graphicx,color}
\usepackage{stmaryrd}

\newcommand{\gmult}{*}

\newcommand{\Fq}{\mathbb{F}_{q}}
\newcommand{\Fpx}[1]{\mathbb{F}_{#1}}

\newcommand{\Psibar}{\overline{\Psi}}
\newcommand{\cardp}[2]{#1 \,\sslash\, #2}
 
\newcommand{\uf}{U_{\!f}}

\def\openone{\leavevmode\hbox{\small1\kern-3.8pt\normalsize1}}
\def\RR{{\rm I\kern-.2emR}}

\def\openone{\leavevmode\hbox{\small1\kern-3.8pt\normalsize1}}
\def\RR{{\rm I\kern-.2emR}}

\providecommand{\ignore}[1]{}

\newcommand{\scalarZ}{0}
\newcommand{\scalarO}{1}
\newcommand{\scalarPlus}{+}

\newcommand{\ip}[2]{\langle #1 ~|~ #2 \rangle}
\newcommand{\usat}{\texttt{UNIQUE-SAT}}
\newcommand{\boolt}{\textsf{Bool}} 

\newcommand{\btrue}{\texttt{\textbf{true}}}
\newcommand{\ff}[1]{\mathbb{F}_{#1}}
\newcommand{\dotprod}{dot product}

\makeatletter
\def\imod#1{\allowbreak\mkern8mu({\operator@font mod}\,\,#1)}
\makeatother

\newcommand{\bitem}{\begin{itemize}}
\newcommand{\eitem}{\end{itemize}}
\newcommand{\benum}{\begin{enumerate}}
\newcommand{\eenum}{\end{enumerate}}
\newcommand{\beq}{\begin{equation}}
\newcommand{\eeq}{\end{equation}}
\newcommand{\beqa}{\begin{eqnarray}}
\newcommand{\eeqa}{\end{eqnarray}}
\newtheorem{definition}{Definition}

\newtheorem{proposition}{Proposition}

\newcommand{\bproof}{\begin{proof}}
\newcommand{\eproof}{\end{proof}}
\newcommand{\bprop}{\begin{proposition}}

\newcommand{\bdef}{\begin{definition}}

\def\round{\mathop{\rm round}\nolimits}

\input{Qcircuit}

\begin{document}
\title{Discrete Quantum Theories}
\author{Andrew J. Hanson${}^{1}$, Gerardo Ortiz${}^{2}$, 
        Amr Sabry${}^{1}$, and Yu-Tsung Tai${}^{3}$}
\affiliation{${}^{1}$School of Informatics and Computing, Indiana University,  Bloomington IN 47405, USA}
\affiliation{${}^{2}$Department of Physics, Indiana University, Bloomington, Indiana 47405, USA}
\affiliation{${}^{3}$Department of Mathematics, Indiana University, Bloomington, Indiana 47405, USA}
\begin{abstract}
  We explore finite-field frameworks for quantum theory and quantum
  computation.  The simplest theory, defined over unrestricted finite fields,
  is unnaturally strong.  A second framework employs only finite fields with
  no solution to $x^2+1=0$, and thus permits an elegant complex
  representation of the extended field by adjoining $i=\sqrt{-1}$.  Quantum
  theories over these fields recover much of the structure of conventional
  quantum theory except for the condition that vanishing inner products arise
  only from null states; unnaturally strong computational power may still
  occur. Finally, we are led to consider one more framework, with further
  restrictions on the finite fields, that recovers a local transitive order
  and a locally-consistent notion of inner product with a new notion of
  \emph{cardinal probability}. In this framework, conventional quantum
  mechanics and quantum computation emerge locally (though not globally) as
  the size of the underlying field increases. Interestingly, the framework
  allows one to choose \emph{separate} finite fields for system description and for
  measurement: the size of the first field quantifies the resources needed to
  describe the system and the size of the second quantifies the resources
  used by the observer.   This resource-based
  perspective potentially provides insights into quantitative measures
  for actual computational power, the complexity of quantum system
  definition and evolution, and the independent question of the cost of
  the measurement process.
 \end{abstract}
\maketitle

\section{Introduction} 

The means by which quantum computing~\cite{qcdef} extends the capacity of
classical computing frameworks deeply involves both the laws of physics and
the mathematical principles of computation.  Richard Feynman and Rolf
Landauer~\cite{RF,RL}, among others, have strongly advocated the careful
study of quantum computation to understand its mechanisms and the source of
its computational features.  Our purpose here is exploit the consequences of
replacing complex continuous numbers by finite complex fields in the quantum
computation framework~\cite{padicqm}; in particular, we show how a number of
subtle properties of quantum computing can be teased apart, step by step, as
we explore the implications of {\it discrete quantum theories\/} in a
systematic fashion.

We observe that the traditional mathematical framework of complex number
fields in quantum mechanics is, in principle, not amenable to numerical
computation with finite resources.  Since theories based on finite fields
are, in principle, always computable with finite resources, a universe that
was in some way a computational engine (a non-trivial philosophical
hypothesis) could actually have a fundamental basis in finite fields, with
conventional quantum mechanics emerging as a limiting case.  This is another
mechanism by which the frameworks we present could conceivably be relevant to
our understanding of the laws of physics.  Specifically, we can quantify the
resources needed for problems of a given complexity by identifying the size
of the required discrete field.  The cost of such resources, clearly exposed
by using discrete fields, is concealed by the properties of real numbers in
conventional quantum computations.

After a review of finite fields in Section~\ref{sec:background}, we proceed
with a sequence of finite-field approaches that lead more and more closely to
the properties of conventional quantum computing. In
Section~\ref{modalquantum}, we examine previously-introduced quantum theories
defined over unrestricted finite fields and show in
Section~\ref{modalquantumcomputing} that this approach leads to theories with
such bizarre powers that they are probably unphysical. Although a version of
quantum theory defined over a two-valued field can express simple algorithms
such as quantum teleportation, it is so weak that it cannot express Deutsch's
algorithm. This quantum theory is, however, also so powerful that it can be
used to solve an unstructured database search of size $N$ using $O(\log(N))$
steps, which outperforms the known asymptotic bound $O(\sqrt{N})$ in
conventional quantum computing.

Next, in Section \ref{discretequantumtheoryI}, we improve on this by showing
that for finite fields of order $p^2$, with the prime $p$ of the form $4 \ell
+ 3$ ($\ell$ a non-negative integer), the complex numbers have extremely
compelling and natural discrete analogs that permit a great many of the
standard requirements of quantum computing to be preserved.  Under suitable
conditions, we have amplitude-based partitions of unity, unitary
transformations, and entanglement, as well as solutions to deterministic
quantum algorithms such as the algorithms of Deutsch, Simon, and
Bernstein-Vazirani~\cite{NCbook,Mermin}, though still with some bothersome
shortcomings.  Because of the modular nature of arithmetic in the finite
complex field, it is not possible to define an inner product in the usual
sense, and we show in Section~\ref{discretequantumcomputingI} that this leads
to excessive computational power for the unstructured database search problem
for certain database sizes.

We are led, in Sections~\ref{discretequantumtheoryIIa}
and~\ref{discretequantumtheoryIIb}, to develop a framework with further
restrictions on~$p$ that \emph{locally\/} recovers the structure and expected
properties of conventional quantum
theory. Section~\ref{discretequantumtheoryIIa} locally recovers the inner
product space and Section~\ref{discretequantumtheoryIIb} locally recovers a
notion of probability. The development in both sections exploits the fact
that longer sequences of {\it ordered\/} numbers appear in the quadratic
residues (numbers with square-roots in the field) as the size of the field
increases.  Discrete quantum computations whose calculations are confined to
numbers in this ordered sequence resemble conventional quantum
computations. The size of the field $p$ plays an important role in describing
the resources needed for the computation as larger problem sizes require a
larger field size to represent all intermediate numerical values. A
significant feature of our framework is that the resources needed for the
measurement process can be separated from the resources needed by the
evolution of the system being modeled. This interplay between the resources
used by the system under study and the resources used for the observation
process is a significant concept that is nonexistent in conventional quantum
computing and is exposed by our careful accounting of resources.

We note that the conventional mathematical framework based on the real
numbers allows one to distinguish states whose measurement outcomes differ by
infinitesimally small probabilities, e.g., $10^{-100}$ vs.~$0$.  In the
proposed framework of discrete quantum computing, the finite size of the
field implies a maximum precision for measurement: a ``small'' field
represents limited resources with which it becomes impossible to distinguish
states whose measurement outcomes differ by an amount less than the
resolution afforded by the field.  It is possible, however, to discriminate
between such states at the cost of moving to a larger field, i.e., by
investing more resources in the measurement process.  We formalize this
approach to measurement using the novel notion of {\it cardinal
  probability\/}, with numerical labels corresponding to ``more probable, the
same, or less probable,'' rather than a percentage-based likelihood measure.
In cardinal probability, relative outcomes are associated with intervals of
ambiguity that get smaller and more precise as the size of the field
increases.

Finally, in Section~\ref{discretequantumcomputingII}, we apply our discrete
quantum theory to the study of two representative algorithms, the
deterministic Deutsch-Jozsa algorithm and the probabilistic Grover
algorithm~\cite{NCbook,Mermin}.  The first algorithm highlights the role
played by the size of the field $p$ in determining the actual resources
required for computation as the number of input bits $n$ increases, a concept
nonexistent in conventional quantum computing.  The second algorithm
highlights, in addition, the dependence of the precision of measurement (via
cardinal probabilities) on the size of the field, another nonexistent concept
in conventional quantum computing.

\section{Fundamentals of Finite Fields}
\label{sec:background}

A field~$\mathbb{F}$ is an algebraic structure consisting of a set of
elements equipped with the operations of addition, subtraction,
multiplication, and division \cite{fieldtheory.ref, numtheory.ref}.  Fields
may contain an infinite or a finite number of elements. The rational
$\mathbb{Q}$, real $\mathbb{R}$, and complex numbers $\mathbb{C}$ are
examples of infinite fields, while the set $\mathbb{F}_3=\{ 0,1,2\}$, under
multiplication and addition modulo 3, is an example of a finite field.

There are two distinguished elements in a field, the addition identity $0$,
and the multiplication identity $1$.  Given the field $\mathbb{F}$, the
closed operations of addition, ``$+$,'' and multiplication, ``$\gmult$,''
satisfy the following set of axioms:
\begin{enumerate}
\item $\mathbb{F}$ is an Abelian group under the addition operation~$+$
  (additive group);
\item The multiplication operation~$\gmult$ is associative and
  commutative. The field has a multiplicative identity and the property that
  every nonzero element has a multiplicative inverse;
\item Distributive laws: For all $a,b,c \in \mathbb{F}$
\begin{eqnarray}
a \gmult (b+c) &=& a \gmult b + a \gmult c \\ 
(b+c) \gmult a &=& b \gmult a+ c \gmult a  \ .
\end{eqnarray}
\end{enumerate} 
\noindent From now on, unless specified, we will omit the symbol~$\gmult$
whenever we multiply two elements of a field.

Finite fields of $q$ elements, $\Fq=\{0,\ldots,q-1\}$, will play a special
role in this work.  A simple explicit example is $\mathbb{F}_3$ with the
following addition and multiplication tables:
\[\begin{array}{c|ccc}
+ & 0& 1 & 2\\[.1in] \hline
0 &  0 &  1  & 2 \\
1  &  1& 2 & 0 \\
2  & 2  & 0 & 1
\end{array}    \hspace*{2cm}
\begin{array}{c|ccc}
\gmult  & 0 & 1 & 2 \\[.1in] \hline
0 &  0 &  0  & 0 \\
1  &  0& 1 & 2 \\
2  & 0  & 2 & 1
\end{array} 
\]

The characteristic of a field is the least positive integer~$m$ such that
$m=1+1+1+\cdots+1 = 0$, and if no such $m$ exists we say that the field has
characteristic zero (which is the case for $\mathbb{R}$ for example). It
turns out that if the characteristic is non-zero it must be a prime $p$. For
every prime $p$ and positive integer $r$ there is a finite field $\Fpx{p^r}$
of size $q=p^r$ and characteristic $p$ (Lagrange's theorem), which is unique
up to field isomorphism. The exponent $r$ is known as the \emph{degree} of
the field over its prime subfield~\cite{galois,GT.ref}. If the characteristic
$p$ is an arbitrary prime number, we call the field \emph{unrestricted}.

For every $a \in \Fq$, $a \neq 0$, then $a^{q-1}=1$, implying the Frobenius
endomorphism (also a consequence of Fermat's little theorem) $a^{q}=a$, which
in turn permits us to write the multiplicative inverse of any non-zero
element in the field as $a^{-1}=a^{q-2}$, since $a^{q-2}a=a^{q-1}=1$. Every
subfield of the field $\Fq$, of size $q=p^r$, has~$p^{r'}$ elements with
some~$r'$ dividing~$r$, and for a given $r'$ it is unique.  Notice that a
fundamental difference between finite fields and infinite fields with
characteristic 0 is one of topology: finite fields induce a compact structure
because of their modular arithmetic, permitting {\it wrapping around}, while
that is not the case for fields of characteristic zero. This feature may lead
to fundamental physical consequences.

\section{Modal Quantum Theory}
\label{modalquantum}

Recently, Schumacher and Westmoreland~\cite{modalqm} and Chang et
al.~\cite{galoisqm} defined versions of quantum theory over {\it
  unrestricted\/} finite fields, which they call modal quantum theories or
Galois field quantum theories. Such theories retain several key quantum
characteristics including notions of superposition, interference,
entanglement, and mixed states, along with time evolution using invertible
linear operators, complementarity of incompatible observables, exclusion of
local hidden variable theories, impossibility of cloning quantum states, and
the presence of natural counterparts of quantum information protocols such as
superdense coding and teleportation.  These modal theories are obtained by
collapsing the Hilbert space structure over the field of complex numbers to
that of a vector space over an \emph{unrestricted} finite field. In the
resulting structure, all non-zero vectors represent valid quantum states, and
the evolution of a closed quantum system is described by \emph{arbitrary}
invertible linear maps.

Specifically, consider a one-qubit system with basis vectors $\ket{0}$
and~$\ket{1}$. In conventional quantum theory, there exists an infinite
number of states for a qubit of the form $\alpha_0\ket{0} + \alpha_1\ket{1}$,
with $\alpha_0$ and $\alpha_1$ elements of the underlying field of complex
numbers subject to the normalization condition
$|\alpha_0|^2+|\alpha_1|^2=1$. Moving to a finite field immediately limits
the set of possible states as the coefficients~$\alpha_0$ and~$\alpha_1$ are
now drawn from a finite set. In particular, in the field $\ff{2}=\{0,1\}$ of
booleans, there are exactly four possible vectors: the zero vector, the
vector~$\ket{0}$, the vector $\ket{1}$, and the vector $\ket{0} +
\ket{1}=\ket{+}$.  Since the zero vector is considered non-physical, a
one-qubit system can be in one of only three states.  The dynamics of these
one-qubit states is realized by any invertible linear map, i.e., by any
linear map that is guaranteed never to produce the zero vector from a valid
state. There are exactly 6 such maps: 
\[\begin{array}{c}
X_0 = \begin{pmatrix}
1 & 0 \\
0 & 1 
\end{pmatrix} , 
\qquad\qquad
X_1 = \begin{pmatrix}
0 & 1 \\
1 & 0 
\end{pmatrix} , 
\\ \\
S \!=\! \begin{pmatrix}
1 & 0 \\
1 & 1 
\end{pmatrix} , 
S^\dagger \!=\! \begin{pmatrix}
1 & 1 \\
0 & 1 
\end{pmatrix} , 
D_1 \!=\! \begin{pmatrix}
0 & 1 \\
1 & 1 
\end{pmatrix} , 
D_2 \!=\! \begin{pmatrix}
1 & 1 \\
1 & 0 
\end{pmatrix} .
\end{array}\]
This set of maps is clearly quite impoverished compared to the full
set of one-qubit unitary maps in conventional quantum theory. In
particular, it does not include the Hadamard transformation.  However,
this set also includes non-unitary maps such as $S$ and $S^\dagger$
that are not allowed in conventional quantum computation.

Measurement in the standard basis is fairly straightforward: measuring
$\ket{0}$ or $\ket{1}$ deterministically produces the same state while
measuring $\ket{+}$ nondeterministically produces $\ket{0}$ or $\ket{1}$
with no assigned probability distribution. In other bases, the measurement
process is complicated by the fact that the correspondence between
$\ket{\Psi}$ and its \emph{dual} $\bra{\Psi}$ is basis-dependent, and that the
underlying finite field is necessarily cyclic. For example, in~$\ff{2}$, addition 
$(+)$ and multiplication $(*)$ are modulo 2: $\ip{+}{+} = (\scalarO * \scalarO) 
+ (\scalarO * \scalarO) = \scalarO + \scalarO = \scalarZ$.
Hence, the dual of $\ket{+}$ is not $\bra{+}$ if $\ket{+}$ is part of the
basis.

\section{Modal Quantum Computing}  
\label{modalquantumcomputing}

To understand the computational implications of the modal quantum theory
defined over the field $\ff{2}$ of booleans, we developed a quantum computing
model and established its correspondence to a classical model of logical
programming with a feature that has quantum-like behavior~\cite{qc-rel}. In a
conventional logic program, answers produced by different execution paths are
collected in a sequence with \emph{no} interference. However, in this modal
quantum computing model over $\ff{2}$, these answers may interfere
destructively with one another.

Our computations with this ``toy'' modal quantum theory showed that it
possesses ``supernatural'' computational power. For example, one can solve a
black box version of the \usat\ problem~\cite{wsusat} in a way that
outperforms conventional quantum computing. The classical \usat\ problem
(also known as \texttt{USAT} or \texttt{UNAMBIGUOUS-SAT}) is the problem of
deciding whether a given boolean formula has a satisfying assignment,
assuming that it has at most one such assignment~\cite{complexity}. This
problem is, in a precise sense~\cite{Valiant198685}, just as hard as the
general satisfiability problem and hence all problems in the~NP complexity
class. Our black-box version of the \usat\ problem replaces the boolean
formula with an arbitrary black box. Solutions to this generalized problem
can be used to solve an unstructured database search of size $N$ using
$O(\log{N})$ black box evaluations by binary search on the database. This
algorithm then outperforms the known asymptotic bound $O(\sqrt{N})$ for
unstructured database search in conventional quantum computing.

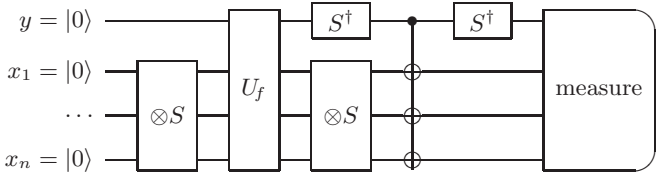
\begin{figure}[t]
\[\begin{array}{@{\!\!}c}
\hspace*{1.35cm} \Qcircuit @C=1.3em @R=.9em {
\lstick{y=\ket{0}}   & \qw                      & \multigate{3}{\uf} &  \gate{~S^\dagger~}         & \ctrl{3} & \gate{~S^\dagger~} & \multimeasureD{3}{\text{measure}} \\
\lstick{x_1=\ket{0}} & \multigate{2}{\otimes S} & \ghost{\uf}        &  \multigate{2}{\otimes S} & \targ    & \qw              & \ghost{\text{measure}}           \\
\lstick{\ldots}      & \ghost{\otimes S}        & \ghost{\uf}        &  \ghost{\otimes S}        & \targ    & \qw              & \ghost{\text{measure}}           \\
\lstick{x_n=\ket{0}} & \ghost{\otimes S}        & \ghost{\uf}        &  \ghost{\otimes S}        & \targ    & \qw              & \ghost{\text{measure}}           
}
\end{array}\]
\caption[]{\label{fig:alg}Circuit for black box \usat\ in modal quantum theory
  over the field $\ff{2}$.  $\uf$ is a Deutsch quantum black
  box~\cite{NCbook} with $ \uf \ket{y}\ket{\overline{x}} ~=~ \ket{y
    \scalarPlus f(\overline{x})}\ket{\overline{x}}$, where $\overline{x}$
  denotes a sequence $x_1, x_2, \ldots, x_n$ of $n$ bits.  For further
  notation see text.}
\end{figure}

We can prove the unreasonable power of the arbitrary-function \usat\ starting
with a classical function $f : \boolt^n \rightarrow \boolt$ that takes $n$
bits and returns at most one \btrue\ result.  Then we can give an algorithm
(see Fig.~\ref{fig:alg}) taking as input such a classical function that
decides, deterministically and in a constant number of black box evaluations,
whether $f$ is satisfiable or not:

\paragraph*{{\rm Case I}: $f$ is unsatisfiable; the measurement
  deterministically produces $\ket{0}\ket{\overline{0}}$.} The state is
initialized to $\ket{0}\ket{\overline{0}}$, with
$\ket{\overline{0}}=\ket{0}\ket{0} \cdots \ket{0}$, i.e., the tensor product
of $n$ $\ket{0}$ states. Applying the map $S$ (defined in the previous
section) to each qubit in the second component of the state produces
$\ket{0}\ket{\overline{+}}$ where $\ket{\overline{+}}$ denotes the sequence
$\ket{+}\ldots\ket{+}$ of length $n$. Applying $\uf$ to the entire state has
no effect since $U_f$ is the identity when $f$ is unsatisfiable. Applying $S$
to each qubit in the second component of the state produces
$\ket{0}\ket{\overline{0}}$. Applying $S^\dagger$ to the first component
leaves the state unchanged.  As the first component of the state is 0,
applying the map $X_0$ (which is the identity) leaves the state unchanged.
Applying $S^\dagger$ to the first component leaves the state
unchanged. Measuring the state will deterministically produce
$\ket{0}\ket{\overline{0}}$.

\paragraph*{{\rm Case II}: $f$ is satisfiable; the measurement produces some
  state other than $\ket{0}\ket{\overline{0}}$.}  Assume the function $f$ is
satisfiable at some input $a_1, a_2, \ldots, a_n$ denoted $\overline{a}$, and
where $\ket{\overline{a}} = \ket{a_1}\ldots\ket{a_n}$. In the second step,
the state becomes $\ket{0}\ket{\overline{+}}$ as above. We can write this
state as $\ket{0}\ket{\overline{a}} + \Sigma_{\overline{x}\neq \overline{a}}
\ket{0}\ket{\overline{x}}$. Applying $U_f$ produces
$\ket{1}\ket{\overline{a}} + \Sigma_{\overline{x}\neq\overline{a}}
\ket{0}\ket{\overline{x}}$. We can rewrite this state as
$\ket{+}\ket{\overline{a}} + \Sigma_{\overline{x}} \ket{0}\ket{\overline{x}}
=\ket{+}\ket{\overline{a}} + \ket{0}\ket{\overline{+}}$, where the summation
is now over all vectors (notice that $\ket{0}\ket{\overline{a}} +
\ket{0}\ket{\overline{a}}$ is the zero vector).  Applying $S$ to each qubit
in the second component produces $\ket{+}\ket{\overline{S(a)}} +
\ket{0}\ket{\overline{0}}$. Applying $S^\dagger$ to the first component
produces: $\ket{1}\ket{\overline{S(a)}} + \ket{0}\ket{\overline{0}}$.
Applying $X_b$, where $b$ is the first component of the state, to each qubit
in the second component produces $\ket{1}\ket{\overline{{\sf not}(S(a))}} +
\ket{0}\ket{\overline{0}}$. Applying $S^\dagger$ to the first component
produces $\ket{+}\ket{\overline{{\sf not}(S(a))}} +
\ket{0}\ket{\overline{0}}$. For the measurement of
$\ket{+}\ket{\overline{{\sf not}(S(a))}} + \ket{0}\ket{\overline{0}}$ to be
guaranteed to never be $\ket{0}\ket{\overline{0}}$, we need to verify that
$\ket{+}\ket{\overline{{\sf not}(S(a))}}$ has one occurrence
$\ket{0}\ket{\overline{0}}$. This can be easily proved as follows. Since each
$a_i$ is either 0 or 1, then each $S(a_i)$ is either $+$ or $1$, and hence
each ${\sf not}(S(a_i))$ is either~$+$ or~$0$. The result follows since any
state with a combination of $+$ and $0$, when expressed in the standard
basis, would consist of a superposition containing the state $\ket{0\ldots}$.



\section{Discrete Quantum Theory (I) } 
\label{discretequantumtheoryI} 

Our next objective is to develop more realistic discrete quantum theory
variants that exclude ``supernatural'' algorithms such as the one presented
above.  Our first such plausible framework~\cite{powerdqc} is based on
complexifiable finite fields. To incorporate complex numbers for quantum
amplitudes, we exploit the fact that the polynomial $x^2+1=0$ is
\emph{irreducible} (has no solution) over a prime field $\ff{p}$ with $p$ odd
if and only if $p$ is of the form~$4\ell+3$, with $\ell$ a non-negative
integer.  In other words, $x^2+1=0$ is irreducible over $\ff{3}, \ff{7},
\ff{11}, \ff{19}, \ldots$.  We achieve our goal by observing that any field
in this family is extensible to a field~$\ff{p^2}$ whose elements can be
viewed as discrete complex numbers with the real and imaginary parts
in~$\ff{p}$. In the field~$\ff{p^2}$, the Frobenius automorphism of an
element $\alpha$ (defined as $\alpha^p$) represents the usual definition of
complex conjugation~\cite{fieldexample}.

The next task is to examine the consequences when we attempt to construct
$d$-dimensional vector spaces over the complexified fields $\ff{p^2}$ \cite{geom1}. (For
readability, instead of writing column vectors, we will often use the vector
notation $\ket{\Psi} = (\alpha_0~\alpha_1~\ldots~\alpha_{d-1})^T$ and
$\ket{\Phi} = (\beta_0~\beta_1~\ldots~\beta_{d-1})^T$, where $(.)^T$ is the
transpose of the {\it row} vector $(.)$.) It can be
shown~\cite{HermitianProd} that, given two vectors
\begin{equation}
\ket{\Psi} = \sum_{i=0}^{d-1} ~\alpha_i \ket{i} \ , \ 
\ket{\Phi} = \sum_{i=0}^{d-1} ~\beta_i \ket{i} , 
\end{equation}
with scalars $\alpha_i$ and $\beta_i$ drawn from the field elements, and
orthonormal basis $\{\ket{i}\}$, the Hermitian \dotprod\ is always reducible
to the form \vspace{-0.1in}
\begin{equation}
\ip{\Phi}{\Psi} = \sum_{i=0}^{d-1} ~\beta_i^p~\alpha_i^{\;} \ .
\label{innerprod}
\vspace{-0.1in} 
\end{equation}

This product satisfies conditions A and B below, but not~C, because in a
finite field, addition can ``wrap around,'' making the concepts of positive
and negative meaningless and allowing the sum of non-zero elements to be zero:
\begin{itemize}
\item[A.] $\ip{\Phi}{\Psi}$ is the complex conjugate of $\ip{\Psi}{\Phi}$; 
\vspace*{-0.25cm}
\item[B.] $\ip{\Phi}{\Psi}$ is conjugate linear in its first argument and
linear in its second argument;
\vspace*{-0.25cm}
\item[C.] $\ip{\Psi}{\Psi}$ is always non-negative and is equal to~0 only
if $\ket{\Psi}$ is the zero vector. 
\end{itemize}

With just conditions A and B, it is possible to recover unitary operators,
and thus recover much of the relevant structure of Hilbert spaces over the
field of complex numbers. The failure of condition C, however, plays havoc
with the traditional notions of ordered probabilities as well as the
geometric notions of ordered distances and angles, whose lengths and cosines,
respectively, are normally expressed using the inner product~\cite{geom}. In
a separate development, we explore the geometry of these finite fields and
define a discrete version of the Hopf fibration extending the Bloch sphere to
$n$-qubits, as well as determining discrete measures for the relative sizes of
the entangled, maximally entangled, and unentangled discrete
states~\cite{geom1}.

\section{Discrete Quantum Computing (I)}
\label{discretequantumcomputingI}

Given a complexified finite field $\ff{p^2}$ and its Hermitian \dotprod\
(Eq.~\eqref{innerprod}) much of the structure of conventional quantum
computing can be recovered.  For example, the smallest field~$\ff{3^2}$ is
already rich enough to express the standard Deutsch-Jozsa~\cite{NCbook}
algorithm, which requires only normalized versions of vectors or matrices
with the scalars $0$, $1$, and $-1$. Similarly, other deterministic quantum
algorithms (algorithms for which we may determine the outcome with
certainty), such as Simon's and Bernstein-Vazirani, perform as
desired~\cite{simon}. Algorithms such as Grover's search will not work in the
usual way because we lack (the notion of) ordered angles and probability in
general.

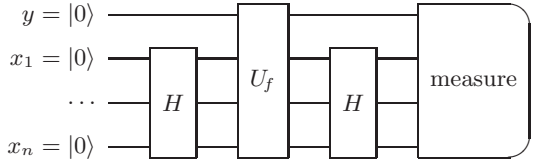
\begin{figure}[t]
\[\begin{array}{c}
\hspace*{1.5cm} \Qcircuit @C=1.7em @R=.9em {
\lstick{y=\ket{0}}   & \qw              & \multigate{3}{\uf} &  \qw              & \multimeasureD{3}{\text{measure}} \\
\lstick{x_1=\ket{0}} & \multigate{2}{H} & \ghost{\uf}        &  \multigate{2}{H} & \ghost{\text{measure}}           \\
\lstick{\ldots}      & \ghost{H}        & \ghost{\uf}        &  \ghost{H}        & \ghost{\text{measure}}           \\
\lstick{x_n=\ket{0}} & \ghost{H}        & \ghost{\uf}        &  \ghost{H}        & \ghost{\text{measure}}           
}
\end{array}\]
\caption[]{Circuit for black box \usat\ in discrete quantum computing.}
\label{fig:dbsearch}
\end{figure}

It is possible, in some situations, to exploit the cyclic behavior of the
field to creatively cancel probability amplitudes and solve problems with
what again appears to be ``supernatural'' efficiency.  We illustrate this
behavior with the algorithm in Fig.~\ref{fig:dbsearch}, which is a variant of
the one in Fig.~\ref{fig:alg}.  Unlike the modal quantum theory algorithm,
the new algorithm does not always succeed deterministically using a constant
number of black box evaluations.  We can, however, show that supernatural
behavior occurs if the characteristic $p$ of the field divides $2^N - 1$.
For a database of fixed size $N$, matching the conditions becomes less likely
as the size of the field increases.  Nevertheless, for a {\it given\/} field,
it is always possible to expand any database with dummy records to satisfy
the divisibility property.  Physically, we are taking advantage of additional
interference processes that happen because of the possibility of ``wrapping
around'' due to modular arithmetic. We do not know, in general, whether this
version of discrete quantum computing actually enables the rapid solution of
NP-complete problems.

\section{Discrete Quantum Theory (II): Inner Product Space} 
\label{discretequantumtheoryIIa}

We next discuss an approach using finite complexifiable fields that
conditionally resolves the inner product condition (C), which is violated by
the theory just presented.  A possible path is suggested by the work of
Reisler and Smith~\cite{finitefieldmemo}.  The general idea is that while the
cyclic properties of arithmetic in finite fields make it impossible to
\emph{globally} obtain the desired properties of the conventional Hilbert
space inner product, it {\it is\/} possible to recover them \emph{locally},
thereby restoring, with some restrictions, all the usual properties of the
inner product needed for conventional quantum mechanics and conventional
quantum computing.  As the size of the discrete field becomes large, the size
of the locally valid computational framework grows as well, leading to the
\emph{effective emergence of conventional quantum theory}.  We next briefly
outline such a context for local orderable subspaces of a finite field, and
introduce an improvement on the original method~\cite{finitefieldmemo}
suggested by recent number theory resources~\cite{yutsungBook}.

Let us first note that the range of the quadratic map, $\{x^2 \!  \ {\rm
  modulo } \ p ~|~ x \in \ff{p}\}$, is always one-half of the non-zero
elements of $\ff{p}$, and is the set of elements with square roots in the
field.  This is the set of \emph{quadratic residues}, and the complementary
set (the other half of the non-zero field elements) is the set of
\emph{quadratic non-residues}.  For example, in $\ff{7}$, the elements
$\{1,2,4\}$ are considered positive as they have the square roots $\{1,3,2\}$
respectively; the remaining elements $\{3,5,6\}$ do not have square roots in
the field. What is interesting is that if we have an uninterrupted sequence
of numbers that are all quadratic residues, then we can define a {\it
  transitive order\/}, with $a>c$ if $a>b$ and $b>c$, provided $a-b$, $b-c$,
and $a-c$ are all quadratic residues.

As a concrete example, consider a finite field in which the sequential
elements $0,1,2,3,\ldots,k-1$ are all quadratic residues (including 0).  Then
any sequence of odd length~$k$ and centered around an arbitrary $x \in
\ff{p}$, i.e., $S_x(k)=x - (k-1)/2, \ldots, x-2, x-1, x, x+1, x+2, \ldots,
x+(k-1)/2$, is {\it transitively ordered}. Indeed, we have $(x+1)-x=1$ which
is a quadratic residue and hence $(x+1) > x$. Similarly, $x - (x-1) = 1$ and
hence $x > (x-1)$. Also $(x+1) - (x-1) = 2$ which is a quadratic residue and
hence $(x+1) > (x-1)$.  Clearly this process may be continued to show that
the sequence $S_x(k)$ is transitively ordered. We can construct examples
using the sequence A000229 in the encyclopedia of integer sequences
\cite{yutsungBook,localorder}.  The $n$th element of that sequence (which
must be prime) is the least number such that the $n$th prime is the
\emph{least} quadratic non-residue for the given element. The first few
elements of this sequence are listed in the top row of Table~\ref{pseq}. The
next row lists the number $k$ of transitively ordered consecutive elements in
that field, and $\pi(k)$ in the bottom row is the prime counting function
(the number of primes up to $k$):

\begin{table}[ht]
\begin{eqnarray}
\hspace*{-0.5cm}\begin{tabular}{c|ccccccccccc}
$p$ & 3 & 7 & 23 & 71 & 311 & 479 & 1559 & 5711 & 10559 & 18191 & \ldots \\
\hline
$k$ & {\bf 2} & {\bf 3} & {\bf 5} & {\bf 7} & {\bf 11} & {\bf 13} &
{\bf 17} & {\bf 19} & {\bf 23} & {\bf 29} & \ldots  \\ \hline
$\pi(k)$ & 1 & 2 & 3 & 4 & 5 & 6 & 7 & 8 & 9 & 10 & \ldots  
\end{tabular} \nonumber
\end{eqnarray}
\caption{Number $k$ of transitively ordered elements for a given field  $\ff{p}$.}
\label{pseq}
\end{table}

As an example, consider the field $\ff{23}$. Looking at the squares of the
numbers $\ff{23}=\{0, \ldots, 22\}$ modulo 23, we find the 2-centered
uninterrupted sequence $S_2(5)=\{0,1,2,3,4\}$, followed by $5$, which is both
the smallest quadratic non-residue and the size of the uninterrupted sequence
of quadratic residues (including 0) of interest.  In particular, it is
possible to construct a total order for the elements $S_0(5)=\{-2,-1,0,1,2\}$
in the fields $\ff{23}$, $\ff{71}$, $\ff{311}$, etc., but not in the smaller
fields $\ff{3}$ and $\ff{7}$.



Given a $d$-dimensional vector space over $\ff{p^2}$ where~$p$ is one of the
primes above, it is possible to define a \emph{region} over which an inner
product and norm can be identified. Let the length of the sequence of
quadratic residues be $k$. The region of interest includes all vectors
$\ket{\Psi}= \sum_{i=0}^{d-1} \alpha_i \ket{i} =
(\alpha_0~\alpha_1~\ldots~\alpha_{d-1})^T$, for which $d < p - \frac{k-1}{2}$
and each $\alpha_i$ satisfies
\begin{eqnarray}
\label{eq:region}
d\,|\alpha_{i}|^{2}=d\,(a_{i}^{2}+b_{i}^{2})\leq\frac{k-1}{2} \ ,
\end{eqnarray}
with $a_i$ and $b_i$  drawn from the set $S_0(k)$. 
Consider, for example, $\ff{311^2}$ ($p=311$, $k=11$).  We find the following
situation in which we can trade off the dimension $d$ of the vector space
against the range of probability amplitudes available for each $\alpha_i$:
\begin{table}[ht]
\begin{tabular}{c|l} 
    & {\sf allowed probability amplitudes} $F^d\left(k\right)$ \\
\hline
$d=1$ & $F^1(11)=$\\
&$\{0,\pm1,\pm 2, \pm i, \pm 2 i, (\pm 1 \pm i), \newline (\pm 1 \pm 2 i), (\pm 2 \pm i) \}$ \\
$d=2$ & $F^2(11)=\{0,\pm1, \pm i,  (\pm 1 \pm i) \}$ \\
$d=3$ & $F^3(11)=\{0,\pm1, \pm i \}$ \\
$d=4$ & $F^4(11)=\{0,\pm1, \pm i \}$ \\
$d=5$ & $F^5(11)=\{0,\pm1, \pm i \}$ \\
$d\geq 6$ & $F^d(11)=\{ 0 \}$ 
\end{tabular}
\caption{Allowed probability amplitudes for different vector space 
dimensions $d$ and $k=11$.}
\label{table_allowed}
\end{table}

We can now verify, by using Table \ref{table_allowed}, that for any vector
$\ket{\Psi}$ in the selected region the value of $\ip{\Psi}{\Psi}$ is $\geq
0$ and vanishes precisely when $\ket{\Psi}$ is the zero vector. Thus, in the
selected region, condition (C) is established.  Although the set of vectors
defined over that region is not closed under addition, and hence the set is
not a vector subspace, we can still have a theory by restricting our
computations.  In other words, \emph{as long as our computation remains
  within the selected region}, we may pretend to have an inner product space.
The salient properties of conventional quantum mechanics emerge, but the
price to be paid is that the state space is no longer a vector space. This is
basically a rigorous formulation of Schwinger's intuition~\cite{schwinger}.

Readers with backgrounds in computer science or numerical analysis will
notice, significantly, that this model for discrete quantum computing is
reminiscent of practical computing with a classic microprocessor having only
integer arithmetic and a limited word length.  We cannot perform a division
having a fractional result at all, since there are no fractional
representations; we do have the basic constants zero and one, as well as
positive and negative numbers, but multiplications or additions producing
results outside the integer range wrap around modulo the word length and
typically yield nonsense.  This implies that, for the local discrete model,
we must accept an operational world view that {\it has no awareness of the
  value of $p$\/}, and depends on having set up in advance an environment
with a field size, analogous to the word size of a microprocessor, that
happily processes {\it any\/} calculation we are prepared to perform.  This
is the key step, though it may seem strange because we are accustomed to
arithmetic with real numbers: we list the calculations that must be performed
in our theory, discover an {\it adequate size of the processor word\/} ---
implying a possibly ridiculously large value of $p$ chosen as described above
--- and from that point on, we calculate necessarily valid values within that
processor, never referring in any way to $p$ itself in the sequel.


\section{Discrete Quantum Theory (II): Cardinal Probability} 
\label{discretequantumtheoryIIb}

The final issue that must be addressed in the discrete theory put forward in
Section~\ref{discretequantumtheoryIIa} concerns measurement. To recap, within
the theory, states are $d$-dimensional vectors with complex discrete-valued
amplitudes drawn from a totally-ordered range, $F^d(k)$, in the underlying
finite field. These states possess, by construction, absolute squares having
values in the positive integers, and squared projections on the bases in the
non-negative integers, all in the ordered range of Eq.~\eqref{eq:region}, and
hence potentially produce probabilities that can be ordered. We start by
applying the measurement framework of conventional quantum computing to these
states; we then systematically expose and isolate the parts that rely on
infinite precision real numbers and replace them by finite approximations.
Our point is that, although the mathematical framework of conventional
quantum mechanics relies on infinite precision probabilities, it is
impossible in practice to measure exact equality of real numbers --- we can
only achieve an approximation within measurement accuracy. Significantly,
when we use finite fields, this measurement accuracy will be encoded in the
size of the finite field used for measurements.


\subsection{Theory}

In conventional quantum theory, given an observable~${\cal O}$ with
eigenvalues $\lambda_i$, $i=0,\cdots,d-1$, and orthonormal eigenvectors
$\ket{i}$ (i.e., ${\cal O}\,\ket{i} = \lambda_i \ket{i}$), the probability of
measuring the (non-degenerate) eigenvalue $\lambda_i$ in a system
characterized by the state $\ket{\Psi}$ is given by:
\begin{eqnarray}
P_{\Psi}(\lambda_i) \equiv P_{\Psi}(i) &=& \frac{|\ip{i}{\Psi}|^2}{\ip{\Psi}{\Psi}}=\frac{|\alpha_i|^2}{\ip{\Psi}{\Psi}}\  ,
\label{probOfEll.eq}
\end{eqnarray}
where $\ket{\Psi} = (\alpha_0~\alpha_1~\ldots~\alpha_{d-1})^T$ in the
eigenbasis of $\cal O$, that is the measurement basis.  Hereafter, we will simplify 
by calling $P_{\Psi}(i)$ the probability of measuring $\lambda_i$.

The fundamental property of conventional quantum theory is that a complete
set of states such as $\{\ket{i}\}$ induces a partition of unity in the ({\it
  real-valued\/}) probabilities, so that
\begin{eqnarray}
\sum_{i=0}^{d-1} P_{\Psi}(i) &=& 1 \ ,
\end{eqnarray}
and, more importantly for our treatment, for any given system, there is a
{\it precise ordering\/} of the set $\{ P_{\Psi}(i) \}$.  In general, this
ordering can be expressed as a sequence of equalities and inequalities of the
following form,
\begin{eqnarray}
\begin{array}{ccccccccc}
 P_{\Psi}(a) & \preceq & P_{\Psi}(b) & \preceq &\cdots&\preceq&
 P_{\Psi}(y) & \preceq & P_{\Psi}(z) \ ,
\end{array}
\label{cqt-order.eq}
\end{eqnarray}
where we adopt the symbol ``$\preceq$'' to denote {\it either\/} equal ($=$)
{\it or\/} less-than ($<$), but {\it never\/} less-than-or-equal ($\leq$).
We observe that {\it in practical measurements\/}, these formal properties
are meaningless, since, statistically, a poor measurement could reverse the
apparent order of the strictly increasing theoretical inequalities; more
significantly, distinguishing {\it formally equal\/} probabilities from a
($>$) or ($<$) ordering is impossible with an observer that has only finite
resources.

We now show that, while formal achievement of the conventional quantum
probability ordering of Eq.~\eqref{cqt-order.eq} is not possible in a world
with finite resources modeled by our discrete quantum theory, we can define a
context for the definition of probabilities, {\it cardinal probability\/},
that is consistent with the just-noted properties of probability measurement
in conventional quantum theory.  That is, in a theory with cardinal
probability, {\it inequalities\/} in the conventional probability relations
Eq.~\eqref{cqt-order.eq} can be preserved with appropriate resources (in the
form of a sufficiently large choice of the field), while {\it equalities\/}
cannot be guaranteed in the theory, and in fact can be represented as {\it
  inequalities of any order\/}.  The {\it set\/} of discrete theories obeying
these properties is defined as a single equivalence class of cardinal
probability theories.

In order to study the explicit properties of a discrete theory, we examine
states of the form
\[\ket{\Psi_{m}} = (\alpha_0^m~\alpha^m_1~\ldots~\alpha^m_{d-1})^T ,\] 
where the coefficients must be discrete complex numbers~$\alpha^m_i$ in the
field representing the resources needed by the computation, and the label $m$
is the ``starting value'' of the discrete norm-squared,
\[ 
m = \ip{\Psi_m}{\Psi_m} = \sum_{i=0}^{d-1} | \alpha^m_{i}|^{2} \ . 
\] 
(We drop the superscript $m$ on the coefficients when there is no ambiguity.)

One might hope to construct a probability object corresponding exactly to the
conventional quantum theory by finding a common factor that eliminated the
diverse denominators $\sqrt{m}$ that would be used to normalize all the
states to unity in the conventional theory.  This would require rescaling
  \[ \frac{\ip{\Psi_{m_1}}{\Psi_{m_1}}}{m_1} =
      \frac{\ip{\Psi_{m_2}}{\Psi_{m_2}}}{m_2}   = 1 \ ,\]
for any two  vectors $\ket{\Psi_{m_1}}$ and $\ket{\Psi_{m_2}}$, to the form
\begin{equation}
\ip{\Psi_{m_1}}{\Psi_{m_1}} \! \prod_{i\neq 1}m_{i} \!  =\!
  \ip{\Psi_{m_2}}{\Psi_{m_2}}\! \prod_{i\neq 2}m_{i}\!  =\!
 \prod_{i}m_{i} \equiv \mu \ .
\label{muNorm.eq}
\end{equation}
Can we succeed in imposing such a restriction?  To determine the answer, let
us define an {\it integer-valued\/} normalization for each of a set of states
we wish to compare: let
 \[ 
\ket{\Psi_{m}} \rightarrow \ket{\Psibar_{m}} = \nu_{m} \ket{\Psi_{m}} = 
  (x_{m}+ i y_{m}) \ket{\Psi_{m}} \ ,
\]
where, from Eq.~\eqref{muNorm.eq}, we would {\it like\/} to have 
\[ \ip{\Psibar_{m}}{\Psibar_{m}} = \prod_{i} m_{i} = \mu \ , \]
or
\[ \ket{\Psibar_{m}} = \left(\prod_{m_{i}\neq m} \sqrt{m_{i}}\right)
\ket{\Psi_{m}}  \]
for {\it any\/} value of $m$. 

({\it Remark:\/} We will take $y_{m}=0 $ in general; replacing a square by a
sum of squares in the norm-squared value of $\ket{\Psibar_{m}}$ adds a few
more cases with exact solutions, but fails to make a difference in the
general case.)

Then we need to show that
\begin{equation}
\ip{\Psibar_{m}}{\Psibar_{m}} = m \, (x_{m})^{2} = \mu
\label{sameMu.eq} 
\end{equation}
either does or does not have a solution for all $m$ in any chosen set
$\{\ket{\Psibar_{m}}\}$.  The resulting condition is obviously
\begin{equation}
m_{1} \, (x_{1})^{2} =  m_{2}  \, (x_{2})^{2}=  m_{3} \, (x_{3})^{2}= \cdots \ . 
\label{mu-eqns.eq}
\end{equation}
Since every $m$ is a sum of at least four squares, even for a single qubit
state, by Lagrange's four-square theorem there is some complexified integer
field that can produce any arbitrary integer as the value of $m$.  Assume
$m_{1}=2$ and $m_{2} = 3$.  Then $x_{2}/x_{1} = \sqrt{2/3}$; but there are no
integer values of $(x_{1},\,x_{2})$ that can satisfy that equation, so it is
impossible in the integer domain to satisfy Eq.~\eqref{mu-eqns.eq} in
general.

This no-go theorem leads us inevitably to consider a {\it set\/} of values of
$\mu_{m}= m \, (x_m)^{2}$ that defines {\it approximate\/} norm-squared
values that are close enough so that the values of the {\it scaled\/}
probabilities based on the set $\{\ket{\Psibar_{m}}\}$ obey the {\it cardinal
  order\/} of Eq.~\eqref{cqt-order.eq} with the following variant of
Eq.~\eqref{probOfEll.eq}:
\[ 
\bar{P}_{\Psibar_{m}}(i) = (x_m)^{2} \, | \alpha_{i}|^{2}  \ .
\]
We notice that $m$ itself does not appear, and that, since each $\alpha_{i}
\rightarrow x_{m} \alpha_{i}$, the original expression is now re-weighted by
$(x_m)^{2}$.  The important point is now that {\it as long as the
  inequalities of Eq.~\eqref{cqt-order.eq} are preserved, and the violation
  of exact equalities does not violate the inequalities\/}, we have a valid
instance of a cardinal probability theory.

\begin{table*}[th!]
\begin{equation} \begin{array}{|c||c|c|c|c|} \hline
{\sf \mbox{Norm$^2$}} = m& \ip{\Psi_{1}}{\Psi_{1}} = 1 & \ip{\Psi_{2}}{\Psi_{2}}=2 &
  \ip{\Psi_{3}}{\Psi_{3}} = 3& \ip{\Psi_{4}}{\Psi_{4}}=4 \\ \hline
\mbox{\sf Prob.~of~} \lambda_0& |\ip{0}{\Psi_{1}}|^2 = \cardp{1}{1} &
|\ip{0}{\Psi_{2}}|^2=\cardp{1}{2} & 
  |\ip{0}{\Psi_{3}}|^2 = \cardp{1}{3}& |\ip{0}{\Psi_{4}}|^2=\cardp{2}{4} \\ \hline
{\sf Prob.~of~} \lambda_1& |\ip{1}{\Psi_{1}}|^2 =\cardp{0}{1} &
|\ip{1}{\Psi_{2}}|^2=\cardp{1}{2} & 
  |\ip{1}{\Psi_{3}}|^2 = \cardp{2}{3}& |\ip{1}{\Psi_{4}}|^2=\cardp{2}{4}  \\ \hline
\end{array} \nonumber
\end{equation}
\caption[]{\label{f11-proj.fig}Norms-squared and probabilities for 
one-qubit states $\ket{\Psi_{m}}$ in $F^2(11)$.}
\end{table*}

Since the ordering requirements typically refer to {\it sets\/} of
comparisons, possibly with different states, we introduce the notation
\begin{equation}
\bar{P}(i) = \cardp{\{ \bar{P}_{\Psibar_{m}}(i) \}} {\{\mu_m\}}
\end{equation}
that expands to
\[ \cardp{\{ (x_{m_1})^{2} | \alpha^{m_{1}}_{i}|^{2},\, (x_{m_2})^{2} |
  \alpha^{m_{2}}_{i}|^{2},\, \cdots \}} {\{\mu_{m_1},\,\mu_{m_2},\, \cdots \}} .
\]
This expression represents a \textit{realization} of the set of cardinal
probabilities $\bar{P}_{\Psibar_{m}}(i)$ with respect to the approximate
normalizations $\mu_{m} = \ip{\Psibar_{m}}{\Psibar_{m}} = m \, (x_m)^{2}$.

The set $\{\mu_{m}\}$ represents the scale with respect to which we are going
to compare cardinal probabilities of states $\{\ket{\Psibar_{m}}\}$ during
the measurement process.  The number of resources required by the observer
corresponds precisely to the characteristic of the field used to define the
scale via the set $\{\mu_{m}\}$.  One can intuitively picture the elements of
$\{\mu_{m}\}$ as a set of {\it rulers\/} that are ``equal'' to within a
certain precision; to get more precision, one needs to buy a more expensive
set of rulers.  Alternatively, one can visualize the precision of the rulers
to be controlled by a set of interactive dials or sliders, with the precision
(as well as the cost of the resources) increasing progressively as the values
are increased.

\subsection{Scale Determination}

We begin with some simple examples of scale determination. Let $p=311$,
$k=11$, and $d=2$.  The permitted range of coefficients is $S_0(11) =
\{-5,\ldots,-1,0,1,\ldots,5\}$; given the dimension $d=2$, the allowed
probability amplitude coefficients are $F^2(11)=\{0,\pm1, \pm i, (\pm 1 \pm
i) \}$ (see Table \ref{table_allowed} above). Consider a single state
$\ket{\Psi_3} = 1 \ket{0} + (1+i) \ket{1}$. In this case there is no need to
scale the state, i.e, we can take $x_m=1$ and calculate $|1|^2=1$,
$|(1+i)|^2=2$ and the norm-squared $\ip{\Psi_3}{\Psi_3} = 3$ (which is in the
allowed range).  The probability of measuring $\lambda_0$ is $\cardp{1}{3}$
and that of measuring $\lambda_1$ is $\cardp{2}{3}$. These results \emph{can}
be used to infer that the probability of measuring $\lambda_1$ is greater
than the probability of measuring $\lambda_0$ but they \emph{cannot} be used
to conclude that the former event is exactly twice as likely as the second.

Now let us consider a more interesting example that involves several
representative one-qubit states,
\[\begin{array}{rcl}
\ket{\Psi_1} &=& 1 \ket{0}  \\
\ket{\Psi_2} &=& 1 \ket{0} + 1 \ket{1} \\
\ket{\Psi_3} &=& 1 \ket{0} + (1+i) \ket{1} \\
\ket{\Psi_4} &=& (1-i) \ket{0} + (1+i) \ket{1} \ ,
\end{array}\]
as explicit examples of each $m$. (There are of course many
equivalent vectors representing the same physical state, a miniature
local version of the traditional Bloch sphere mapping \cite{geom1}.)
Table~\ref{f11-proj.fig} presents the bare analogs of norms-squared and
probabilities for the $\ket{\Psi_{m}}$ representing the properties of the
four unique norms, $m=1,\ 2,\ 3,\ {\mathrm{and}~} 4$.  (In larger fields,
these numbers do not necessarily form a sequence.)

\bigskip

We will introduce a deterministic construction to identify approximate
choices for $\{\mu_{1},\mu_{2},\mu_{3},\mu_{4}\}$ in a moment.  But first let
us give a clear heuristic example of the nature of the problem and the
process by which we can converge towards solutions. In
Table~\ref{approxMu.tbl}, we show two guesses for the values of
$\{x_1,x_2,x_3,x_4\}$.  The first is extremely simple, but the numbers do not
quite have enough power to avoid a conflict with the required order
corresponding to the real-valued probabilities $P_{\Psi_m}(0)=(1, 1/2, 1/3,
1/2)$ and $P_{\Psi_m}(1)=(0, 1/2, 2/3, 1/2)$.  The second choice, still
constructed from integers that are quite small, achieves the required
ordering and is our first example of an instance of a cardinal probability
system for $\{\ket{\Psi_1},\,\ket{\Psi_2},\, \ket{\Psi_3},\, \ket{\Psi_4}
\}$.

\begin{table*}[th!]
\vspace{-.1in}
\begin{equation*} \begin{array}{|c||c||c|c|c||c||c|c|c|} \hline
&\multicolumn{4}{c||}{\sf Failing\  Choice}&\multicolumn{4}{c|}{\sf Successful
\   Choice} \\ \hline
\textrm{Actual}~\{P_{\Psi_m}(0),P_{\Psi_m}(1)\} &m & x &\mu=m x^2 & \{\bar{P}_{\Psibar_m}(0),\bar{P}_{\Psibar_m}(1)\}
&m & x &\mu=m x^2 & \{\bar{P}_{\Psibar_m}(0),\bar{P}_{\Psibar_{m}}(1)\}\rule[-.60em]{0em}{1em} \\ \hline
 \{1,0\}& 1& 4& 16& \{16,0\} & 1 & 16 & 256 &  \{256,0\} \\
\{1/2,1/2\}& 2& 3& 18& \{9,9\} & 2 & 12 & 288 &  \{144,144\} \\
\{1/3,2/3\}& 3& 2& 12& \{4,8\} & 3 & 9 & 243 &  \{81,162\} \\
\{1/2,1/2\}& 4& 2& 16& \{8,8\}  & 4 & 8 & 256 &  \{128,128\} 
 \\ \hline 
\end{array}
\vspace{-.1in}
\end{equation*}
\caption[]{\label{approxMu.tbl} A failing choice (left) and a
  successful choice (right) for the rescaling of the
  system $\{\ket{\Psi_1},\,\ket{\Psi_2},\, \ket{\Psi_3},\,
  \ket{\Psi_4} \}$ to realize a cardinal probability system consistent 
  with conventional quantum mechanical probabilities.}
\end{table*}

\bigskip

To extend this heuristic framework toward a deterministic computation, we now
propose specific criteria to select the set of normalizations
$\{\mu_1,\mu_2,\mu_3,\mu_4\}$ with respect to which we can compare cardinal
probabilities. The method relies on introducing the notion of {\it square
  root\/} of a number drawn from a finite field.  In conventional quantum
computing, it is possible to re-weight the four states above so that all have
a norm-squared of 24 as follows:
\[\begin{array}{rcl}
\ket{\Psi_1} &=& 2\sqrt{6} \ (1 \ket{0})  \\
\ket{\Psi_2} &=& 2\sqrt{3} \ (1 \ket{0} + 1 \ket{1}) \\
\ket{\Psi_3} &=& 2\sqrt{2} \ (1 \ket{0} + (1+i) \ket{1}) \\
\ket{\Psi_4} &=& \sqrt{6} \ ((1-i) \ket{0} + (1+i) \ket{1}) \ .
\end{array}\]
However, it is impossible to achieve this re-weighting precisely in a discrete
theory because the square roots cannot be calculated exactly in finite fields.
We can, however, produce successively more accurate approximations 
of square roots with bigger and bigger fields
using a prescription suggested by Reisler and Smith~\cite{finitefieldmemo}.


We denote the approximate square root of $m>0$ in a finite field $\ff{p}$ by
$\sqrt[']{m}$. This approximate square root is calculated by taking the usual
square root of the smallest element in the ordered range $S_0(k)$ that is
greater than~$m$ and that is a quadratic residue.  For example, in a field
with more than 8 positive ordered elements, $S_0(k\geq 19)$, we have:
\[\begin{array}{rclcl}
\sqrt[']{2} &=& \sqrt{4} &=& 2 \\
\sqrt[']{3} &=& \sqrt{4} &=& 2 \\
\sqrt[']{6} &=& \sqrt{9} &=& 3  \ .
\end{array}\]
Even though these approximations are crude, they can be used to re-weight the
vectors above to get probabilities $\bar{P}_{\Psibar_{m}}(i)$ whose
relationships approximate the ideal mathematical (but uncomputable using 
finite resources) probabilities. In more detail, the re-weighted vectors
become:
\[\begin{array}{rcl}
\ket{\overline{\Psi}_1} &=& 6~(1 \ket{0}) \\
\ket{\overline{\Psi}_2} &=& 4~(1 \ket{0} + 1 \ket{1}) \\
\ket{\overline{\Psi}_3} &=& 4~(1 \ket{0} + (1+i) \ket{1}) \\
\ket{\overline{\Psi}_4} &=& 3~((1-i) \ket{0} + (1+i) \ket{1}) \ ,
\end{array}\]
with $\{\mu_m\}= \{36,\,32,\,48,\,36\}$, 
and the probabilities become:
\[\begin{array}{rcl@{\qquad\qquad}rcl}
\bar{P}_{\Psibar_{1}}(0) &=& 36   & \bar{P}_{\Psibar_{1}}(1) &=& 0 \\
\bar{P}_{\Psibar_{2}}(0) &=& 16   & \bar{P}_{\Psibar_{2}}(1)  &=& 16 \\
\bar{P}_{\Psibar_{3}}(0) &=& 16   & \bar{P}_{\Psibar_{3}}(1)  &=& 32 \\
\bar{P}_{\Psibar_{4}}(0) &=& 18   & \bar{P}_{\Psibar_{4}}(1)  &=& 18 \ ,
\end{array}\]
which we express as 
\[\begin{array}{rcl}
\bar{P}(0) &=& \cardp{\{ 36, 16, 16, 18 \}}{\{ 36, 32, 48, 36 \}} \\
\bar{P}(1) &=& \cardp{\{ 0, 16, 32, 18 \}}{\{ 36, 32, 48, 36 \}} \ .
\end{array}\]
In comparison with the exact probabilities, we see that
$\bar{P}_{\Psibar_{3}}(0)$ and $\bar{P}_{\Psibar_{2}}(0)$ collapse to a
single value and $\bar{P}_{\Psibar_{4}}(0)$ is approximated in a way that
makes it larger than $\bar{P}_{\Psibar_{2}}(1) $. If we only concern
ourselves with how the actual probabilities are related by the $\preceq$
relation, then our approximation is adequate.

If we desire an even more accurate approximation, we can proceed as follows:
We choose a larger field for measurement in which the ordered ranged is
scaled by 100 so that the square roots get one additional digit of
precision. Specifically, in a field with more than 625 positive ordered
elements, we have
\[\begin{array}{rclcl}
\sqrt[']{200} &=& \sqrt{225} &=& 15 \\
\sqrt[']{300} &=& \sqrt{324} &=& 18 \\
\sqrt[']{600} &=& \sqrt{625} &=& 25 \ ,
\end{array}\]
giving a better approximation of the square roots (scaled by 10). 
Using these approximations, the four vectors become:
\[\begin{array}{rcl}
\ket{\Psibar_1} &=& 50~(1 \ket{0}) \\
\ket{\Psibar_2} &=& 36~(1 \ket{0} + 1 \ket{1}) \\
\ket{\Psibar_3} &=& 30~(1 \ket{0} + (1+i) \ket{1}) \\
\ket{\Psibar_4} &=& 25~((1-i) \ket{0} + (1+i) \ket{1})
\end{array}\]
with $\{\mu_m\}=\{2500,\,2592,\,2700,\,2500\}$, 
and the probabilities become:
\[\begin{array}{rcl@{\qquad\qquad}rcl}
\bar{P}_{\Psibar_{1}}(0) &=& 2500   & \bar{P}_{\Psibar_{1}}(1) &=& 0 \\
\bar{P}_{\Psibar_{2}}(0) &=& 1296   & \bar{P}_{\Psibar_{2}}(1)  &=& 1296 \\
\bar{P}_{\Psibar_{3}}(0) &=& 900   & \bar{P}_{\Psibar_{3}}(1)  &=& 1800 \\
\bar{P}_{\Psibar_{4}}(0) &=& 1250   & \bar{P}_{\Psibar_{4}}(1)  &=& 1250 \  .
\end{array}\]
In comparison with the exact probabilities, we see that the increase in
precision has reestablished the distinction between $\bar{P}_{\Psibar_{3}}(0)$ and
$\bar{P}_{\Psibar_{2}}(0)$. The two probabilities 
$\bar{P}_{\Psibar_{4}}(0)$ and $\bar{P}_{\Psibar_{2}}(1)$ are now
relatively closer but they are still, however, not equal. A moment's
reflection shows that these two values can \emph{never} be equal as
$(\sqrt[']{2})^2$ can never be precisely $2$ no matter how many digits of the
actual $\sqrt{2}$ we maintain. 

\section{Discrete Quantum Computing (II)}  
\label{discretequantumcomputingII}

We now examine two particularly important types of examples within the
discrete theory of the previous section: the first is the deterministic
Deutsch-Jozsa algorithm~\cite{NCbook,Mermin}, which determines the balanced
or unbalanced nature of an unknown function with a single measurement step
($O(1)$), and the second is the (normally) probabilistic Grover
algorithm~\cite{NCbook,Mermin,Grover}, determining the result of an
unstructured search in $O(\sqrt{N})$ time.  In the following, we use $k$ to
denote the upper bound of the ordered range of integers needed to perform a
given calculation; this in turn is assumed to be implemented using a choice
of a finite prime number $p$ that supports calculation in the range of~$k$.

\subsection{Discrete Deutsch-Jozsa Algorithm: Deterministic}

To examine the Deutsch-Jozsa algorithm in the discrete theory of the previous
section, we assume we are given a classical function
$f:\boolt^{n}\rightarrow\boolt$, and are told that $f$ is either constant or
balanced~\cite{NCbook,Mermin}. The algorithm is expressed in a space of
dimension $d=2^{n+1}$: it begins with the $n+1$ qubit state
$\ket{1}\ket{\overline{0}}$ where the overline denotes a sequence of
length~$n$.  A straightforward calculation~\cite{NCbook} shows that
the final state is~\cite{HadamardNote} 
\[
\sum_{\overline{z}\in\left\{ 0,1\right\} ^{n}}\sum_{\overline{x}\in\left\{ 0,1\right\}^{n}}
\left(-1\right)^{f\left(\overline{x}\right)+\overline{x}\cdot\overline{z}}\left(\ket{0}\ket{\overline{z}}-\ket{1}\ket{\overline{z}}\right) \ ,
\]
and that its norm-squared is $2^{n+1}$. To make sure that the algorithm
works properly, we note that all the probability amplitudes involved in the
calculation are in the range $-2^n,\ldots,2^n$ and therefore, by
Eq.~\eqref{eq:region}, we get the following constraint on the size of the
ordered region in the finite field:
\[
2^{n+1}\left(2^{n}\right)^{2}\leq\frac{k-1}{2}
   \;\;\Leftrightarrow \;\; k \geq 2^{3n+2} + 1 \ .
\]

Now we need to choose a prime number $p$ that supports calculation in
the range of $k$.  Assume that $k$ is the least prime satisfying $k
\geq 2^{3n+2} + 1$, and let $p$ be the $\pi(k)$th element of the
sequence A000229 \cite{yutsungBook}.  We argue that no prime less than
this value of $p$ can support calculation in the ordered range of $k$, and
that this $p$ is sufficient to support such calculation.  In
particular, since $k$ is the least quadratic non-residue of $p$, every
number less than $k$ is a quadratic residue, and thus
$0,1,2,3,\ldots,2^{3n+2}$ are all quadratic residues.  Hence the
numbers $-2^n,\ldots,2^n$ are all inside the ordered range $S_0(k)$. 
 On the other hand, if we choose any prime smaller than $p$, there is
 a quadratic non-residue smaller than $k$, and we also know that the
 least quadratic non-residue is a prime \cite{numtheory.ref}.  Thus, there is a
quadratic non-residue in $0,1,2,3,\ldots,2^{3n+2}$ , and therefore,
for this smaller $p$,  there would be a
number in $-2^n,\ldots,2^n$ that is not in the ordered range $S_0(k)$.

When $f$ is constant, the cardinal probability of measuring
$\ket{0}\ket{\overline{0}}$ or $\ket{1}\ket{\overline{0}}$ is
$\cardp{\left(2^{n}\right)^{2}+\left(2^{n}\right)^{2}=2^{2n+1}}{2^{2n+1}}$;
i.e., the cardinal probability of measuring any other state
is~$\cardp{0}{2^{2n+1}}$. When $f$ is balanced, the cardinal probability of
measuring $\ket{0}\ket{\overline{0}}$ or $\ket{1}\ket{\overline{0}}$ is
$\cardp{0}{2^{2n+1}}$. Therefore, if we find that the post-measurement
state is either $\ket{0}\ket{\overline{0}}$
or $\ket{1}\ket{\overline{0}}$, we know $f$ is constant; otherwise, $f$ is
balanced.

For a single qubit Deutsch problem, the absolute maximum probability
amplitude is $2$ and $d=2^{1+1}=4$, so we want to have
\begin{eqnarray*}
k & \geq & 2^{3\cdot1+2}+1=2^{5}+1=33 \ .
\end{eqnarray*}
The least prime satisfying the above condition is $k=37$, and thus
\begin{eqnarray*}
\pi\left(37\right) & = & 12\\
p & = & 422231 \ ,
\end{eqnarray*}
where the prime counting function $\pi(k)$ is taken from the
extended elements in  Table \ref{table_long}.
\begin{table}[ht]
\begin{eqnarray} 
\label{pseq2}
\hspace*{0.0cm}
\begin{tabular}{c|ccccccccc}
 \ensuremath{p} &  \ensuremath{\ldots} &  422231  &  \ensuremath{\ldots} &  196265095009  &  \ensuremath{\ldots} &   &  \ensuremath{\ldots} &   &  \ensuremath{\ldots}\\
\hline \ensuremath{k} &  \ensuremath{\ldots} &  {\bf 37}  &  \ensuremath{\ldots} &  {\bf 131}  &  \ensuremath{\ldots} &  {\bf 257}  &  \ensuremath{\ldots} &  {\bf 32771}  &  \ensuremath{\ldots}\\
\hline \ensuremath{\pi(k)} &  \ensuremath{\ldots} &  12  &  \ensuremath{\ldots} &  32  &  \ensuremath{\ldots} &  55  &  \ensuremath{\ldots} &  3513  &  \ensuremath{\ldots}
\end{tabular}
\  \nonumber
\end{eqnarray}
\caption{Extension of transitively ordered elements. }
\label{table_long}
\end{table}

For the 2-qubit Deutsch-Jozsa, the computation is already quite
challenging. Now the absolute maximum probability amplitude is $4$ and
$d=2^{2+1}=8$, so we need
\begin{eqnarray*}
k & \geq & 2^{3\cdot2+2}+1=2^{8}+1=257 \ .
\end{eqnarray*}
Because 257 is a prime, we can pick
\begin{eqnarray*}
k&=&257\\
\pi\left(257\right) & = & 55  \ .
\end{eqnarray*}
The actual value of $p$ is already outside the range of the published
tables.

These examples illustrate that the value of $p$ plays an essential role: its
size grows with the numerical range of the intermediate and final results of
the algorithms being implemented.  Therefore, we naturally recover a
deterministic measure of the intrinsic resources required for a given level
of complexity; this measure is normally completely hidden in computations
with real numbers, and explicitly exposing it is one of the significant
achievements of our discrete field analysis of quantum computation.  This
solves the conundrum that the conventional Deutsch-Jozsa algorithm
mysteriously continues to work for larger and larger input functions without
any apparent increase in resources.  Our analysis of this problem reveals
that as the size of the input increases, it is necessary to increase the size
of $p$ and hence the size of the underlying available numeric coefficients.
This observation does not fully explain the power of quantum computing over
classical computing, but at least it explains that some of the power of
quantum computing depends on increasingly larger precision in the underlying
field of numbers.

\subsection{Discrete Grover Search: Nondeterministic}

As an example of how to apply our cardinal probability framework to a
nondeterministic algorithm, consider the $N\times N$ ``diffusion'' and
``phase rotation'' matrices for searching an unstructured database of
size~$N=2^{n}$ using Grover's algorithm~\cite{Grover}:

\vspace{-0.1in} 
\begin{eqnarray*}
D & = &
\begin{pmatrix}
1 - \frac{N}{2} & 1 & 1 & \ldots & 1 \\
1 & 1 - \frac{N}{2} & 1 & \ldots & 1 \\
1 & 1 & 1 - \frac{N}{2} & \ldots & 1 \\
\vdots & \vdots & \vdots & \vdots & \vdots \\
1 & 1 & 1 & \ldots & 1 - \frac{N}{2}
\end{pmatrix} ,\\
R & = &
\begin{pmatrix}
-1 & 0 & 0 & \ldots & 0 \\
0 & 1 & 0 & \ldots & 0 \\
0 & 0 & 1 & \ldots & 0 \\
\vdots & \vdots & \vdots & \vdots & \vdots \\
0 & 0 & 0 & \ldots & 1
\end{pmatrix} \ ,
\vspace{-0.1in} 
\end{eqnarray*}
where we have eliminated, in matrix $D$, the scaling factor $2/N$ to enforce
the requirement that all matrix coefficients in our framework are
integer-valued.  Note that we have chosen the ``marked'' element in matrix
$R$ to be in the first position. In the standard algorithm, the
transformation $DR$ is repeated $j$ times, where
\[ j =
\round\left(\frac{\pi}{4\arccos\sqrt{1-\frac{1}{N}}}-\frac{1}{2}\right)
\approx \round\left(\frac{\pi}{4}\sqrt{N}\right) \ .
\] 
In our context, we must choose a prime number that is large enough to ensure
that all the numbers that occur during the calculation and after measurement
are within the transitively-ordered subrange.

Let $f$ be the function we want to search, and let $\overline{t}$ be the
target, i.e., $f\left(\overline{x}\right)=1$ if and only if
$\overline{x}=\overline{t}$.  Because the probability amplitudes of
$\ket{\overline{x}}$ are all the same for $\overline{x}\neq\overline{t}$, we
can let $a_{l}$ be the probability amplitude of $\ket{\overline{t}}$, with
$b_{l}$ the probability amplitude of each of the other possibilities, which
are all the same.  We begin at $l=0$ with the information-less state, the
normalization scaled to integer values as usual, which we can write as
\[ 
   \left(\begin{array}{c} a_0 \\ b_0 \\ \vdots\\ b_0 \end{array} \right) = 
   \left(\begin{array}{c} 1 \\ 1 \\ \vdots \\ 1 \end{array} \right) \ . 
\]
Applying the operators $DR$, and denoting by $a_{l}$ and $b_{l}$ the two
unique elements of the $N$-dimensional column vector describing the evolving
process, we find the following recurrence relation for the successive
coefficients:
\begin{eqnarray*}
 a_{0}  & = & 1 \\
 b_{0}  & = & 1 \\
 a_{l+1}  & = & \left(\frac{N}{2} - 1\right)  a_{l}  + (N - 1 ) \, b_{l} \\
 b_{l+1}  & = & (-1 ) \,  a_{l}  + \left(\frac{N}{2} - 1 \right)  b_{l}  \ .
\end{eqnarray*}
We also know $\left|a_{j}\right|>\left|b_{j}\right|$, so we can
estimate an upper bound for the
maximum cardinal probability  as 
\[ \max\left|a_{j}\right|^2 \leq 2\left(\frac{N}{2}\right)^{2j+1} \ .\]
By applying Eq.~\eqref{eq:region} with $d=N=2^n$, we can estimate~$k$ using
\vspace{-0.1in}
\[ 
k\geq 8\left(\frac{N}{2}\right)^{2j+2}+1\ .
\vspace{-0.1in} 
\]

If we pick a prime $k$ satisfying the above condition, then
choosing the $\pi(k)$th prime in the sequence represented by 
Table \ref{pseq} guarantees that every number we need for
the computation is within the transitively ordered range $F^{d}(k)$.

For the 2-qubit Grover search, we have $N=d=4$ and $j=1$, with the maximum
cardinal probability
\[ 
\max\left|a_{j}\right|^2 \leq 2\left(\frac{4}{2}\right)^{2+1}
= 16 \ ,
\]
so we need
\begin{eqnarray*}
k & \geq & 8\left(\frac{4}{2}\right)^{2\cdot1+2}+1=8\cdot2^{4}+1=129 \ .
\end{eqnarray*}
The least prime $k$ satisfying the above condition is $k=131$, and so
\begin{eqnarray*}
\pi\left(131\right) & = & 32 \\
p&=&196265095009 \ .
\end{eqnarray*}

When $p=196265095009$, we assume that  $f\left(\overline{x}\right)=1$
if and only if $\ket{\overline{x}}=\ket{0}\!\ket{0}$, and so the final state is 
$(4,0,\cdots,0)^T$
with norm-squared of $16$. Then, the cardinal probability of obtaining
$\ket{0}\!\ket{0}$ as the post-measurement state is
$\cardp{16}{16}$, and it is $\cardp{0}{16}$ for the rest of the states.

For the 3-qubit Grover search, we have $N=d=8$ and $j=2$, with an
upper bound  $\max\left|a_{j}\right|^2 \leq
2\left(\frac{8}{2}\right)^{4+1} = 2048$ on the cardinal probability.  Thus
\begin{eqnarray*}
 k &\ge& 8 \left(\frac{8}{2}\right)^6 + 1 = 32769 \ .
\end{eqnarray*}
The nearest prime  greater than this number is 32771, so we can pick 
\begin{eqnarray*}
k&=&32771\\
\pi(32771) & = & 3513  \ , 
\end{eqnarray*}
and so if we use the 3513{th} prime, we can implement Grover's algorithm for
a database of size 8.

Continuing with the 3-qubit Grover example, we show how the cardinal
probabilities evolve to single out the target state.  First, assume that
$f\left(\overline{x}\right)=1$ if and only if
$\ket{\overline{x}}=\ket{0}\!\ket{0}\!\ket{0}$.  The initial information-less
8-dimensional state vector evolves under the application of $DR$ as follows:
\[\left(\begin{array}{c}  
1\\
1\\
\vdots\\
1
\end{array}\right)\rightarrow\left(\begin{array}{c}
10\\
2\\
\vdots\\
2
\end{array}\right)\rightarrow\left(\begin{array}{c}
44\\
-4\\
\vdots\\
-4
\end{array}\right) \ .\]
These states have differing norm-squared, so we multiply the first and second
states by $16$ and $4$, respectively, to force them to have the same value of
$2048$.  The now-consistently-normalized states become
\[\left(\begin{array}{c}
16\\
16\\
\vdots\\
16
\end{array}\right)\rightarrow\left(\begin{array}{c}
40\\
8\\
\vdots\\
8
\end{array}\right)\rightarrow\left(\begin{array}{c}
44\\
-4\\
\vdots\\
-4
\end{array}\right) \ . \]
Therefore, the cardinal probabilities of measuring
$\ket{0}\!\ket{0}\!\ket{0}$ in each state are
\[\begin{array}{c@{\qquad}c@{\qquad}c}
\cardp{256}{2048} & \cardp{1600}{2048} & \cardp{1936}{2048} \ ,
\end{array}\]
while the cardinal probabilities of measuring the other states become
\[\begin{array}{c@{\qquad}c@{\qquad}c}
\cardp{256}{2048} & \cardp{64}{2048} & \cardp{16}{2048} \ .
\end{array}\]
We may thus conclude that the cardinal probability of measuring the
satisfying assignment of $f$ increases as we apply the diffusion $D$ and
phase rotation $R$ matrices repeatedly.

Clearly, the required size of $k$ increases systematically with the problem
size, and the corresponding size of the required prime number $p$ defining
the discrete field increases in the fashion illustrated in Tables~\ref{pseq}
and~\ref{table_long}.

\medskip
\section{Conclusions}  

Since conventional quantum theory is defined over uncomputable complex
numbers, it is natural to explore alternative versions of quantum theory
based on finite fields.  Examining the computational and physical
consequences of such computable frameworks can yield new insights into the
power and capacity of quantum computing.  We have described a path through
several variants of discrete quantum theories, starting with unrestricted
discrete fields (modal theories), then advancing to a more reasonable framework
based on complexifiable discrete fields (discrete quantum theory I), which
supports unnaturally efficient deterministic quantum algorithms.  We conclude
with a still more plausible discrete theory (discrete quantum theory II),
from which conventional quantum computing and conventional quantum theory
emerge in a local sense.  Note that as the number of restrictions on the
discrete fields increases, the frequency of possibly unreasonable
efficiency decreases.  As long as we do not perform measurements or the
quantum algorithm is of a deterministic nature, as in Deutsch's problem, we
do not need to invoke any statistical postulates.  This situation
is an exception, since conventional quantum mechanics requires 
probabilistic components describing information extracted by
measurement from the systems being studied.  This 
measurement process is problematic in any discrete quantum theory.  To
resolve the measurement problem in our nondeterministic situations, we have
introduced the notion of cardinal probability.  With this approach,
we see that the issues surrounding
transitively-ordered probability, intrinsically troublesome in quantum
theory for discrete fields, show signs of being resolvable locally.
Interestingly, our framework allows us to define distinct finite fields for system
description and for measurement.  These finite fields distinguish the
resources needed to describe the system from the resources used by the
observer.  Additional work is in progress on the interaction between the
geometrical properties of finite fields and discrete quantum computing, and
we hope to be able to make more definitive statements about probability
measures based on the properties of discrete geometry.  Our investigation
leaves open the question of whether conventional quantum mechanics is
physical, or whether perhaps extremely large discrete quantum theories that
contain only computable numbers are at the heart of our physical universe.



\begin{thebibliography}{99}
\bibitem{qcdef} We use the phrase ``conventional quantum theory'' where
  necessary to distinguish the usual quantum theory and quantum computing
  paradigm using (continuous) complex numbers from discrete quantum theory.
  Alternative terminology in the literature includes ``actual,''
  ``standard,'' and ``ordinary'' quantum theory.
\bibitem{galois}
Fields $\ff{q}$ where $q$ is a power of a prime $p$, i.e.,
$q=p^r$, are  known as Galois fields.
\bibitem{fieldexample}
The simplest example is the field~$\ff{3^2}$ with 9 elements,
\[
0, 1, -1, i, 1+i, -1+i, -i, 1-i, -1-i \ .
\]
(These are all the complex numbers one can form using the integers modulo 3 as
real and imaginary coefficients.)  Similarly, the field $\ff{7^2}$ has 49
elements of the form $\alpha =a+ib$ where $a,b$ are integers in the range $[-3,3]$
and addition and multiplication are modulo 7.
\bibitem{localorder} For computational purposes, this sequence is preferable
  to the one proposed by Reisler and Smith~\cite{finitefieldmemo} because it
  produces smaller primes.  Their work showed that a sufficient condition on
  finite fields to produce sequences of quadratic residues is to further
  constrain the underlying prime numbers to be of the form $8 \Pi_{i=1}^m q_i
  - 1$, where $q_i$ is the $i$th odd prime.  While all such primes are of the
  form $4 \ell + 3$, the set is severely restricted to astronomical numbers
  because the first few such primes are $7, 23, 839, 9239, 2042039, \ldots$.
\bibitem{HadamardNote} Note that the algorithm in
  reference~\cite{NCbook} makes use of the Hadamard matrix.  We have
  eliminated the factor $1/\sqrt{2}$ to ensure that all quantities are
  expressed in terms of integers.  Also notice that the positioning of
  the initial qubit state $\ket{1}$ is reversed from~\cite{NCbook}.
\end{thebibliography}
\end{document}